\documentclass{elsart}

\usepackage[square,comma]{natbib}
\usepackage{graphicx}
\usepackage{pxfonts}
\usepackage{lineno}

\usepackage{amssymb}

\journal{}

\begin{document}

\thispagestyle{empty}
\begin{Large}
\textbf{DEUTSCHES ELEKTRONEN-SYNCHROTRON}

\textbf{\large{Ein Forschungszentrum der Helmholtz-Gemeinschaft}\\}
\end{Large}

DESY 11-081

May 2011

\begin{eqnarray}
\nonumber &&\cr \nonumber && \cr \nonumber &&\cr
\end{eqnarray}
\begin{eqnarray}
\nonumber
\end{eqnarray}
\begin{center}
\begin{Large}
\textbf{The effects of betatron motion on the preservation of FEL
microbunching}
\end{Large}
\begin{eqnarray}
\nonumber &&\cr \nonumber && \cr
\end{eqnarray}

\begin{large}
Gianluca Geloni,
\end{large}
\textsl{\\European XFEL GmbH, Hamburg}
\begin{large}

Vitali Kocharyan and Evgeni Saldin
\end{large}
\textsl{\\Deutsches Elektronen-Synchrotron DESY, Hamburg}
\begin{eqnarray}
\nonumber
\end{eqnarray}
\begin{eqnarray}
\nonumber
\end{eqnarray}
ISSN 0418-9833
\begin{eqnarray}
\nonumber
\end{eqnarray}
\begin{large}
\textbf{NOTKESTRASSE 85 - 22607 HAMBURG}
\end{large}
\end{center}
\clearpage
\newpage

\begin{frontmatter}



\title{The effects of betatron motion on the preservation of FEL microbunching}


\author[XFEL]{Gianluca Geloni\thanksref{corr},}
\thanks[corr]{Corresponding Author. E-mail address: gianluca.geloni@xfel.eu}
\author[DESY]{Vitali Kocharyan}
\author[DESY]{and Evgeni Saldin}

\address[XFEL]{European XFEL GmbH, Hamburg, Germany}
\address[DESY]{Deutsches Elektronen-Synchrotron (DESY), Hamburg,
Germany}

\begin{abstract}
In some options for circular polarization control at X-ray FELs, a
helical radiator is placed a few ten meters distance behind the
baseline undulator. If the microbunch structure induced in the
baseline (planar) undulator can be preserved, intense coherent
radiation is emitted in the helical radiator. The effects of
betatron motion on the preservation of micro bunching in such
in-line schemes should be accounting for. In this paper we present a
comprehensive study of these effects. It is shown that one can work
out an analytical expression for the debunching of an electron beam
moving in a FODO lattice, strictly valid in the asymptote for a FODO
cell much shorter than the betatron function. Further on, numerical
studies can be used to demonstrate that the validity of such
analytical expression goes beyond the above-mentioned asymptote, and
can be used in much more a general context. Finally, a comparison
with Genesis simulations is given.
\end{abstract}

%
%
%
\end{frontmatter}



\section{\label{sec:uno}  Introduction}

The LCLS baseline includes a planar undulator system, which produces
intense linearly polarized light in the wavelength range between
$0.15$ nm and $1.5$ nm \cite{LCLS2}. Several schemes using helical
undulators have been proposed for polarization control at the LCLS
setup \cite{GENG,KUSK,OURC}. The option presented in \cite{OURC},
exploits the microbunching of the planar undulator. After the
baseline undulator, the electron beam is transported along a $40$ m
long straight line by FODO focusing system and subsequently passed
through a helical radiator.  If the microbunch structure of the
bunch can be preserved, intense coherent radiation is emitted in the
helical radiator. The driving idea of this proposal is that the
background linearly-polarized radiation from the baseline undulator
is suppressed by spatial filtering. This operation consists in
letting radiation and electron beam through horizontal and vertical
slits upstream of the helical radiator, where the radiation spot
size is about ten times larger than the electron bunch transverse
size. The effect of betatron motion on the preservation of micro
bunching in such scheme should be accounting for. In fact, the
finite angular divergence of the electron beam, linked with the
betatron function, yields a spread of the longitudinal velocity
leading to microbunching suppression. In \cite{OURC} we estimated
this factor, and concluded that the betatron motion should not
constitute a serious problem in proposed scheme.

In this paper we present a comprehensive study of the effect of
betatron motion on the microbunch preservation. Our paper is based
on the use results of \cite{SVEN}, where it was showed that in the
limit for a small length of a FODO cell with respect to the betatron
function value, the longitudinal velocity of an electron, averaged
over a FODO cell, is constant through the focusing system. Based on
this non-trivial statement, one can work out an analytical
expression for the debunching of an electron beam moving in a FODO
lattice, strictly valid in the asymptote for a short FODO cell.
Further on, numerical studies can be used to demonstrate that the
validity of such analytical expression goes beyond the
above-mentioned asymptote, and can be used in much more a general
context.

The present work is organized as follows. In the following Section
\ref{sec:due} we review the main result in \cite{SVEN}, and we
report the expression for the average longitudinal velocity of an
electron, which depends on the Courant-Snyder invariant of motion.
Based on this result, the analytical expression for the debunching
is calculated. In Section \ref{sec:tre}, the analytical asymptote is
cross-checked with numerical calculations, and its validity is
extended. A comparison with results obtained with the code Genesis
\cite{GENE} is presented. Finally, in Section \ref{sec:conc}, we
come to conclusions.

\section{\label{sec:due}  Analytical study}

The starting point for our analysis is a consideration on the
longitudinal velocity of an electron in a FODO lattice. As has been
shown in \cite{SVEN}, when the length of the FODO cell
$L_\mathrm{FODO}$ is much shorter than the betatron function, the
longitudinal velocity of the electron, averaged over one FODO cell
length, is constant. From a mathematical viewpoint, this result may
be obtained with simple analytical calculations from Eq. (6) and Eq.
(7) of reference \cite{SVEN}. We will begin to consider a 2D motion
on the $x(\mathrm{horizontal})-z(\mathrm{longitudinal})$ plane. The
longitudinal velocity of a certain electron can be written as $v_z
(z)= v [1-\cos(\theta(z))]$, where $v$ is the electron speed, and
$\theta(z) = x'(z)$ is the angle formed at each point of the
electron trajectory with the longitudinal axis. When, as is the case
for ultrarelativistic electrons moving along the $z$ axis, $x'(z)
\ll 1$, one can expand the trigonometric function and obtain $v(z)
\simeq v - v x'(z)^2/2$, where $x'(z)$ is given by Eq. (6) and Eq.
(7) of reference \cite{SVEN}. When the length of FODO cell,
$L_\mathrm{FODO}$, is much shorter then betatron function, the
magnitude of the Twiss parameter $\alpha$ approaches unity, and one
can re-write these two equations approximatively as:

\begin{eqnarray}
&&x'(z) \simeq \sqrt{\frac{I_x}{\beta}} [\cos(\phi) - \sin(\phi)]
~~~\mathrm{when}~z<L_\mathrm{FODO}/2 \cr && x'(z) \simeq
\sqrt{\frac{I_x}{\beta}} [\cos(\phi) + \sin(\phi)]
~~~\mathrm{when}~L_\mathrm{FODO}/2<z<L_\mathrm{FODO}
~.\label{xpfodo}
\end{eqnarray}
Here $I_x$ is the particle Courant-Snyder invariant, while $\beta$
and $\phi$ are the betatron function and the betatron phase
respectively. Averaging $v_z$ over one FODO cell length one obtains:

\begin{eqnarray}
<v_z> = \frac{v}{L_\mathrm{FODO}} \int_0^{L_\mathrm{FODO}} \left[1 -
\frac{x'(z)^2}{2}\right] dz = v\left(1-\frac{I_x}{2\beta}\right)
~,\label{vzave}
\end{eqnarray}

\begin{figure}[tb]
\includegraphics[width=1.0\textwidth]{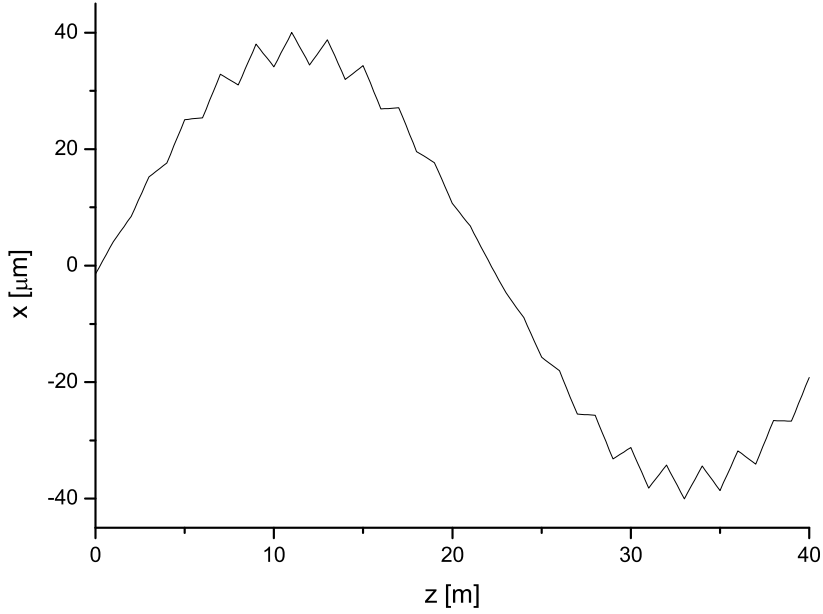}
\caption{Trajectory of an electron within a FODO cell. Here
$L_\mathrm{FODO} = 1.0$ m and $\beta \simeq 10$ m.} \label{betatr}
\end{figure}
which is proportional to the Courant-Snyder invariant $I_x$, and is
independent of $z$. This result looks at first glance surprising. In
fact, a glance of a typical electron trajectory shows, Fig.
\ref{betatr} shows an overall oscillatory trajectory, and one would
expect that the longitudinal velocity at $z=0$, where the electron
trajectory forms an angle with the $z$ axis, should be smaller than,
for example, that at $z \simeq 10$ m, where the trajectory is almost
parallel to the $z$ axis. The physical explanation is that the
longitudinal velocity is decreased at the position for at $z \simeq
10$ m, by the presence of sharp oscillations on the scale of the
FODO cell, which, as demonstrated above, lead to a constant average
longitudinal velocity.

We will now make use of Eq. (\ref{vzave}) to study the effects of
the betatron motion on the preservation of FEL microbunching.
Consider an electron beam carrying an average current $I_0$. Let us
superimpose an initial modulation at a given frequency $\omega$. The
total current can be written as a function of the phase $\psi =
\omega (z/v_z - t)$, with $z$ the longitudinal position, $t$ the
time, $v_z$ the longitudinal velocity as

\begin{eqnarray}
I_1 = I_0 (1 + a_1 \cos \psi) \label{inmod}~.
\end{eqnarray}
Similarly, given the unmodulated longitudinal particle density $n_0
= I_0/(-e v_z)$, $(-e)$ being the electron charge, the longitudinal
particle density $n_1(\psi)$ after the initial modulation is given
by

\begin{eqnarray}
n_1 = n_0 (1 + a_1 \cos \psi) \label{Nnmod}~.
\end{eqnarray}
Here $a_1$ describes the initial bunching. The relation between
$a_1$ and the bunching factor $b_1 \equiv <\exp(i \psi)>$, which
can be often found in literature is given by

\begin{eqnarray}
b_1 = \sum_{k=1}^{N_{\mathrm{ptc}}} \exp(i \psi_k) = \frac{1}{2\pi}
\int_0^{2\pi} d \psi \frac{n_1(\psi)}{n_0} \exp (i\psi) =
\frac{a_1}{2}~,\label{relation}
\end{eqnarray}
where $N_{\mathrm{ptc}}$ is the total number of particles within a
wavelength $\lambda = 2 \pi v_z/\omega$, i.e. $N_{\mathrm{ptc}} =
n_0 \lambda$, and $\psi_k$ is the phase of each particle.

The ratio $L_\mathrm{FODO}/\beta$, with $\beta$ the average betatron
function, is the first relevant parameter of our problem. We will
assume $L_\mathrm{FODO}/\beta \ll 1$, so that Eq. (\ref{vzave}) can
be used for the average longitudinal velocity of an electron. The
phase difference of an electron with Courant-Snyder invariant $I_x$
with respect to one moving on-axis, after a given distance $L$ can
thus be written as $\delta \psi = \Delta v_z L/(2 c \lambdabar)
\simeq - I_x L/(2 \beta \lambdabar)$. Since the rms value for $I_x$
is the geometrical emittance $\varepsilon$  one obtains,
parametrically, $\delta \psi$ is of order $\varepsilon L/(\beta
\lambdabar)$, which is the second and last relevant parameter of our
problem. No debunching is expected for $\varepsilon L/(\beta
\lambdabar) \ll 1$.

Under the accepted approximation $L_\mathrm{FODO}/\beta \ll 1$ it is
possible to derive an analytical expression for the evolution of the
bunching factor along the FODO lattice. In fact, the influence of
the betatron motion alone can be modelled by substituting the phase
$\psi_k$ of each individual electron with $\psi_k + \delta
\psi_k(\theta_x)$ where, as described above,

\begin{eqnarray}
\delta \psi_k =-  \frac{I_{x,k} L}{2
\beta\lambdabar}~.\label{delpsik}
\end{eqnarray}
The bunching factor $b_2$ after the propagation along the FODO cell
can therefore be written as an average over the distribution of
$I_x$, which we call $f(I_x)$, as

\begin{eqnarray}
b_2 && \equiv <\exp[i (\psi_k +  \delta \psi_k)]> =
\sum_{k=1}^{N_\mathrm{ptc}} \exp[i (\psi_k+\delta \psi_k)] \cr && =
\frac{1}{2\pi} \int_0^{2 \pi} d\psi \frac{n_1(\psi)}{n_0}
\int_0^{\infty} d I_x f(I_x)  \exp[i (\psi+\delta
\psi)]~.\label{main}
\end{eqnarray}
Using

\begin{eqnarray}
f(I_x) = \frac{1}{2 \varepsilon} \exp\left[-\frac{I_x
}{2\varepsilon}\right]~, \label{fdist}
\end{eqnarray}
remembering Eq. (\ref{Nnmod}) and  Eq. (\ref{relation}) one obtains
from Eq. (\ref{main}) the following expression for the ratio $\zeta$
between final and initial bunching:

\begin{eqnarray}
\zeta &&\equiv \frac{a_2}{a_1} = \frac{b_2}{b_1} = \frac{1}{ 2
\varepsilon} \int d I_x \exp\left[-\frac{I_x}{2\varepsilon}\right]
\exp\left[-\frac{i I_x L}{2 \beta \lambdabar}\right]~.\label{main2}
\end{eqnarray}
The integral can be calculated analytically yielding the final
result

\begin{eqnarray}
\zeta = \left(1+ i \frac{L \varepsilon}{\beta
\lambdabar}\right)^{-1}~.\label{main3}
\end{eqnarray}

\begin{figure}[tb]
\includegraphics[width=0.5\textwidth]{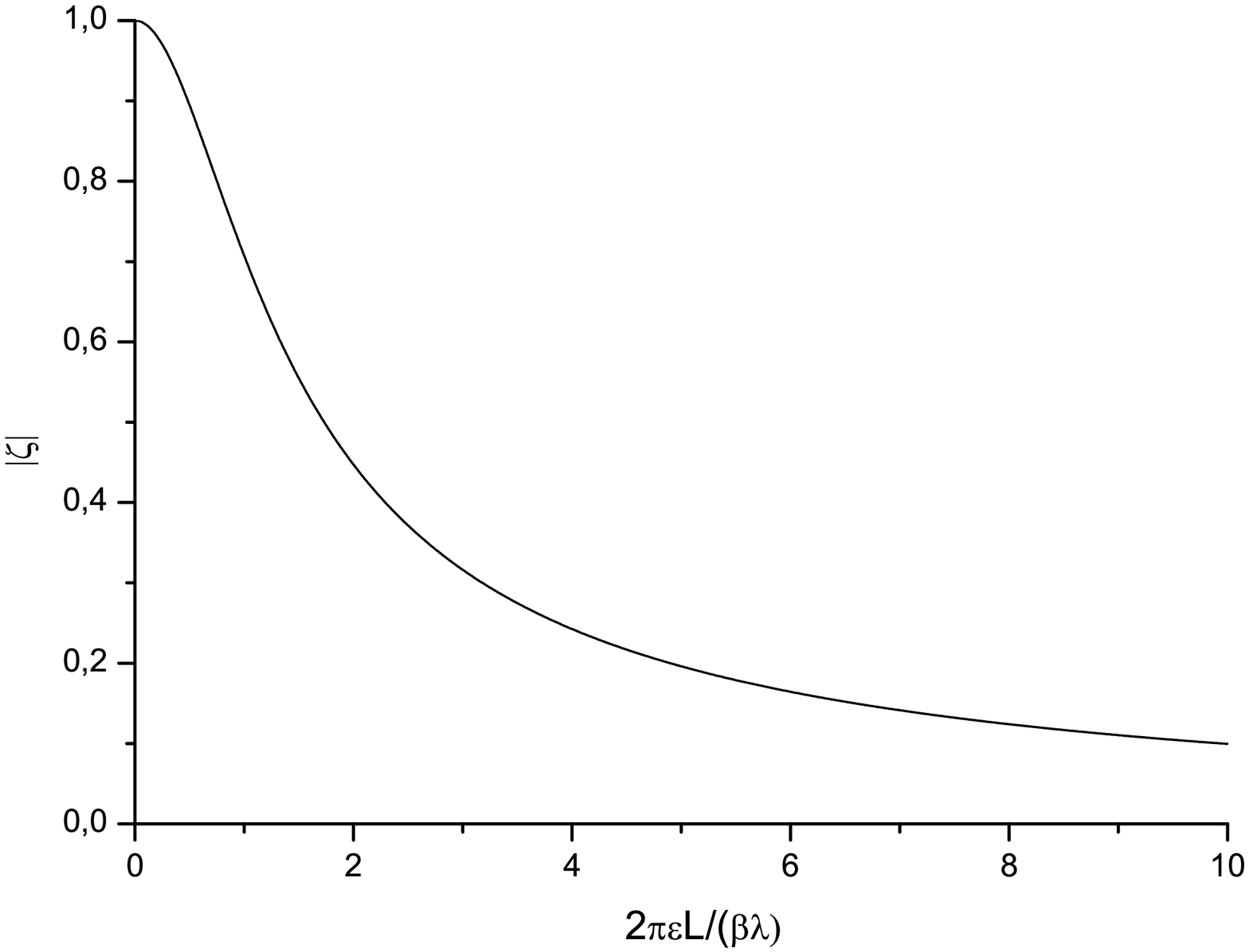}
\includegraphics[width=0.5\textwidth]{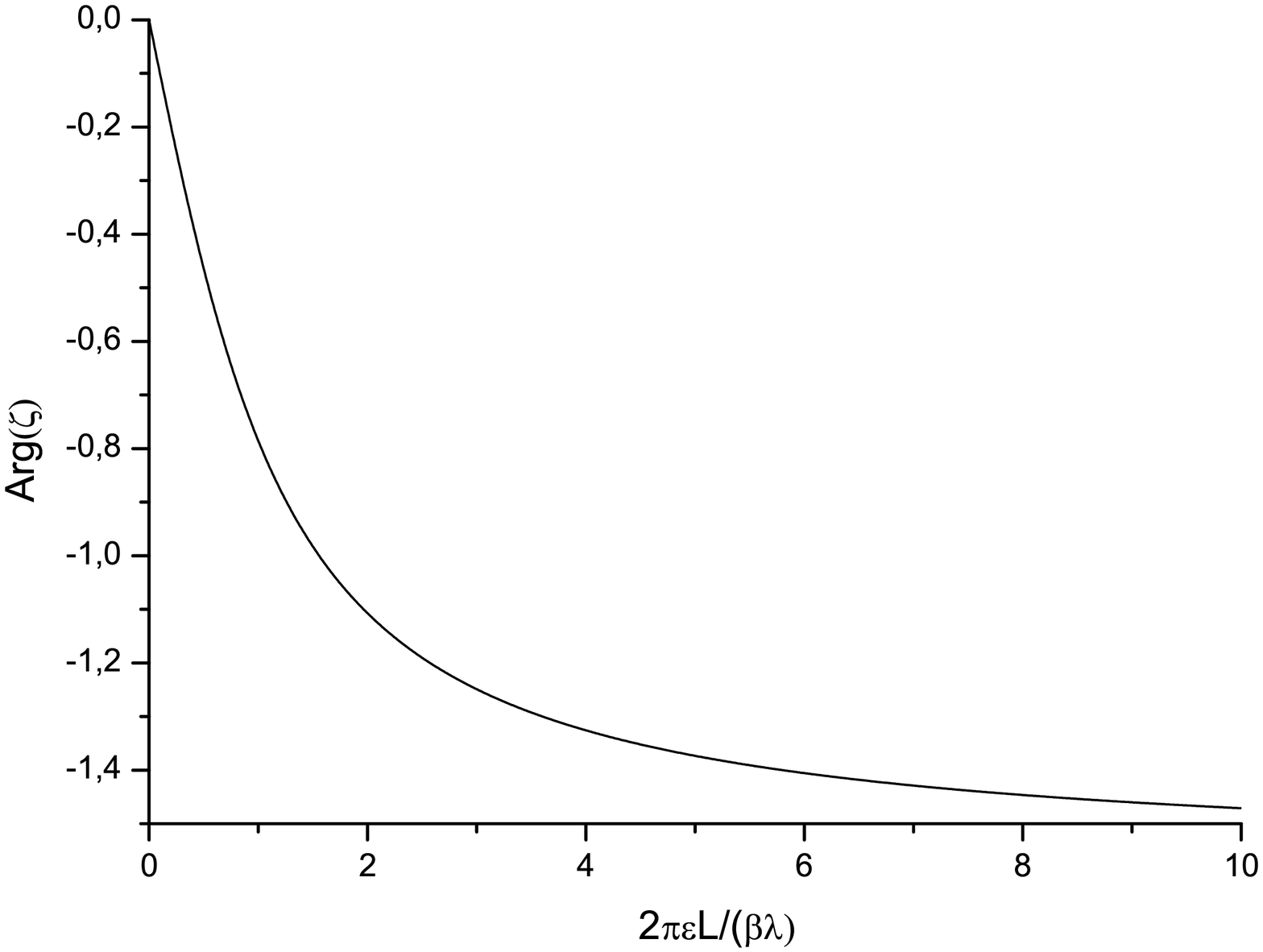}
\caption{Modulus (left plot) and phase (right plot) of $\zeta$ as
a function of the parameter $\varepsilon L/(\beta \lambdabar)$.}
\label{zeta}
\end{figure}
Note that $\zeta$ is not a real number. The physical interpretation
of $|\zeta| = [1+\varepsilon^2 L^2/(\lambdabar^2 \beta^2)]^{-1/2}$
is that of the evolution of the amplitude of the bunching. The
physical interpretation of $\mathrm{Arg}(\zeta)$ is that of the
evolution of the phase of the bunching. In other words, the bunching
evolves in modulus and phase along the FODO lattice. Also, notice
that $\zeta$ is independent of the initial definition of the
bunching, $a_1$ or $b_1$. In both cases, the final bunching can be
found by multiplying the initial bunching by $\zeta$. Modulus and
phase of $\zeta$ are plotted in Fig. \ref{zeta} as a function of the
parameter $\varepsilon L/(\beta \lambdabar)$.

Finally, we note that all considerations have been made for a 2D
motion on the $x-z$ plane. However, generalizing to a 3D motion is
trivial. One needs to account the divergence $y'$, giving an extra
contribution to the phase $\delta \psi$ formally identical as that
for the horizontal direction, once the horizontal beta function is
substituted with the vertical one. One obtains that $\zeta = \zeta_x
\zeta_y$, with $\zeta_{x,y}$ formally identical to Eq.
(\ref{main3}).

\section{\label{sec:tre}  Numerical study}

In order to study the influence of the betatron beating on the
microbunch suppression and to have an idea about the accuracy of our
analytical asymptotic results, we simulated the evolution of the
bunching numerically. In particular, we considered a periodic
lattice composed of drift, focusing element in the thin-lens
approximation, drift, defocusing element. For simplicity, we
considered the motion in the $x-z$ direction, so that Eq.
(\ref{main3}) could be used to compare with debunching calculated
numerically. In the numerical calculations, the magnetic structure
is defined, in terms of quadrupole strength and length of the
drifts. Then, the Twiss parameters at the beginning of the setup are
calculated, and used to generate the horizontal phase-space
distribution of electrons. Each particle is tracked through the
setup in a linear matrix approach, and the trajectory is calculated.
Knowing the trajectory $x_k(z)$ for each electron it is then
straightforward to calculate the curvilinear distance traveled.
Comparison with the length of the setup allows one to recover the
phase difference $\delta \psi_k$ and to calculate $\zeta = <\exp(i
\delta \psi_k)>$.

\begin{figure}[tb]
\includegraphics[width=1.0\textwidth]{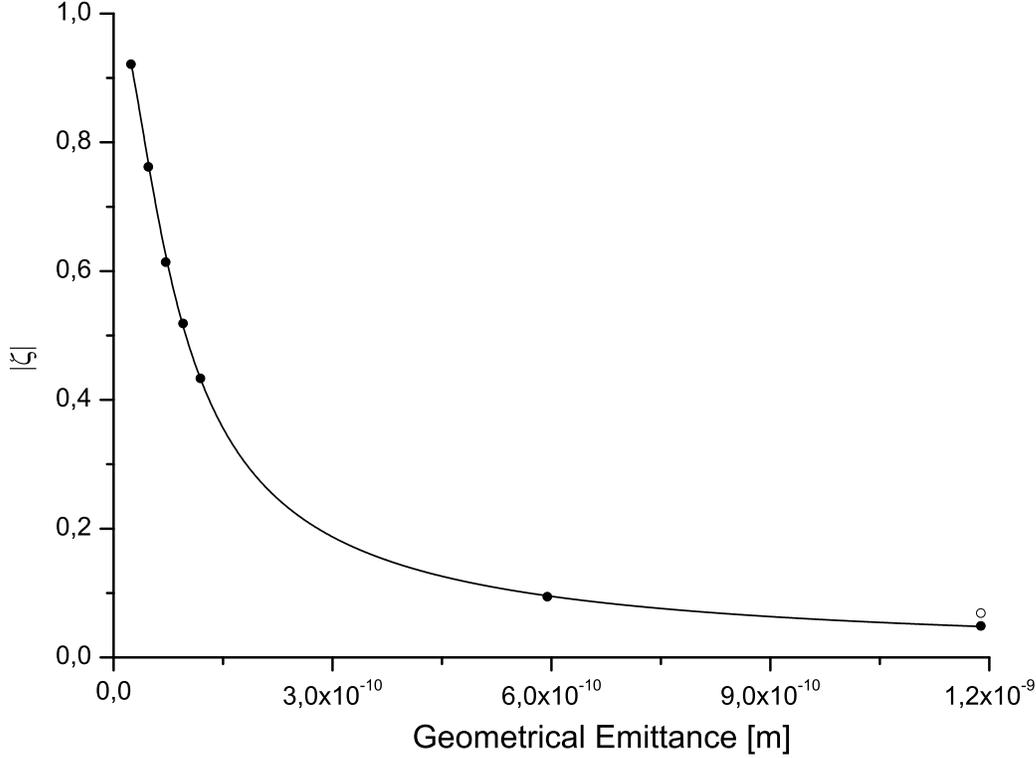}
\caption{Comparison between $|\zeta|$ calculated analytically as a
function of the emittance from Eq. (\ref{main3}) (solid line) and
calculated numerically (black circles) as described in this Section.
Here $\bar{\beta} \simeq 10$m, and $L_\mathrm{FODO} = 1$m. The
modulation wavelength is $\lambda = 1.5$ nm, while the drift length
is $L=40$ m. The number of particles is varied from $10^4$ to
$5\cdot 10^4$ to achieve a good accuracy. As an exemplification, the
white circle corresponds to $10^4$ particles, whereas the black
circle at the same emittance value corresponds to $5 \cdot 10^4$
particles.} \label{emit}
\end{figure}

We began our numerical investigations setting the average betatron
function value $\bar{\beta} \simeq 10$ m, and the length of the cell
$L_\mathrm{FODO} = 1$m. This choice for the length of the FODO cell
allows to consider the asymptote $L_\mathrm{FODO}/\bar{\beta} \ll 1$
nearly satisfied.  The debunching module $|\zeta|$ was calculated
for different emittance value. A comparison with $|\zeta|$
calculated analytically from Eq. (\ref{main3}) is shown in Fig.
\ref{emit}. The number of particles is varied from $10^4$ to $5\cdot
10^4$ to achieve a good accuracy, and is not constant for the
calculated points. For the sake of exemplification, the white circle
in Fig. \ref{emit} corresponds to $10^4$ particles, whereas the
black circle at the same emittance value corresponds to $5 \cdot
10^4$ particles. The good agreement between numerical and analytical
results was to be expected on the basis of the asymptote
$L_\mathrm{FODO}/\bar{\beta} \ll 1$.

\begin{figure}[tb]
\includegraphics[width=1.0\textwidth]{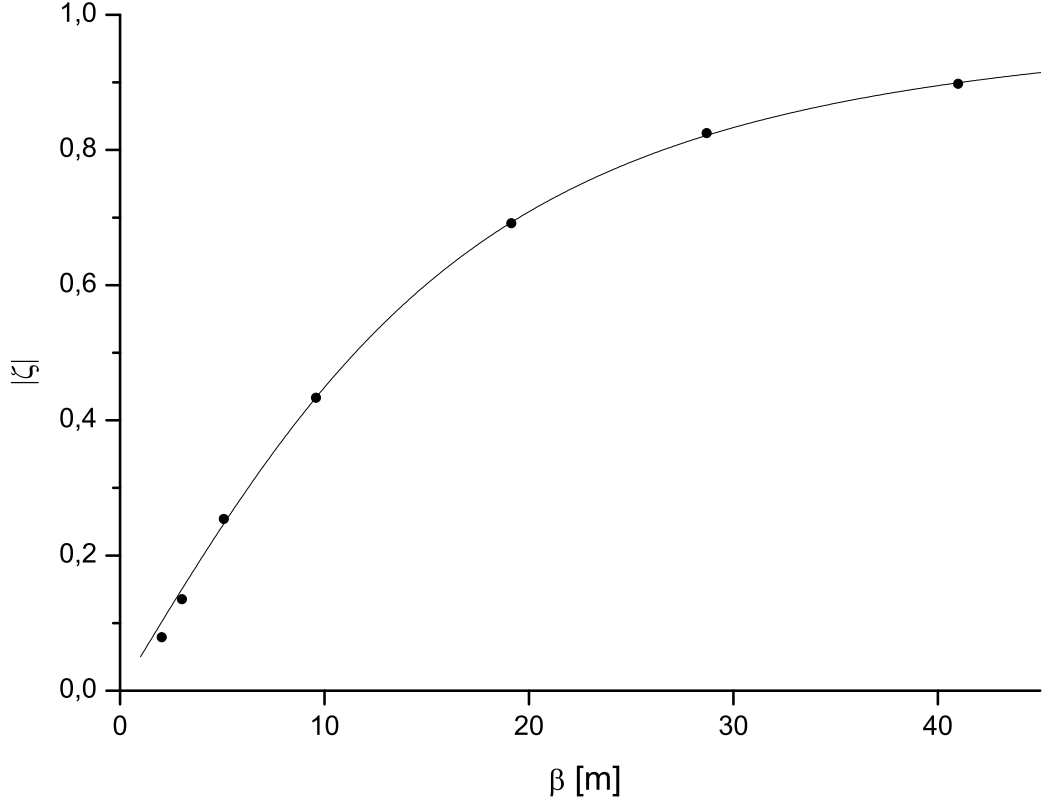}
\caption{Comparison between $|\zeta|$ calculated analytically as a
function of the average betatron function from Eq. (\ref{main3})
(solid line) and calculated numerically (black circles) as described
in this Section. Here $\varepsilon = 1.188 \cdot 10^{-10}$ m, and
$L_\mathrm{FODO} = 1$m. The modulation wavelength is $\lambda = 1.5$
nm, while the drift length is $L=40$ m.} \label{beta}
\end{figure}

An interesting result can be achieved by fixing the emittance (in
our case $\varepsilon = 1.188 \cdot 10^{-10}$ m), and changing
$\bar{\beta}$, keeping the other quantities as in the previous
example. This means that still $L_\mathrm{FODO} = 1$ m. In this way
one can sweep through different values of the parameter
$L_\mathrm{FODO}/\bar{\beta}$. In particular, we calculated
numerically the value of $|\zeta|$ for
$1/40<L_\mathrm{FODO}/\bar{\beta}<1/2$. The result of the comparison
with the asymptotic for $L_\mathrm{FODO}/\bar{\beta} \ll 1$ in Eq.
(\ref{main3}) is shown in Fig. \ref{beta}. A very good agreement can
be seen even for values $L_\mathrm{FODO}/\bar{\beta}$ not much
smaller than unity. Since the average beta function is related to
the betatron phase $\phi$ and to the length of the FODO cell by
$\bar{\beta} = L_\mathrm{FODO} [\cot(\phi/2)+2/3 \tan(\phi/2)]$, the
maximuum value achievable for $L_\mathrm{FODO}/\bar{\beta}$ turns
out to be limited to about $0.61$ for a working focusing system.
From this numerical analysis it follows that Eq. (\ref{main3}) can
be used not only in the asymptotic case for
$L_\mathrm{FODO}/\bar{\beta} \ll 1$ but, with the good accuracy
given in Fig. \ref{beta}, practically in all cases. This is the main
result of this work.

\begin{figure}[tb]
\includegraphics[width=1.0\textwidth]{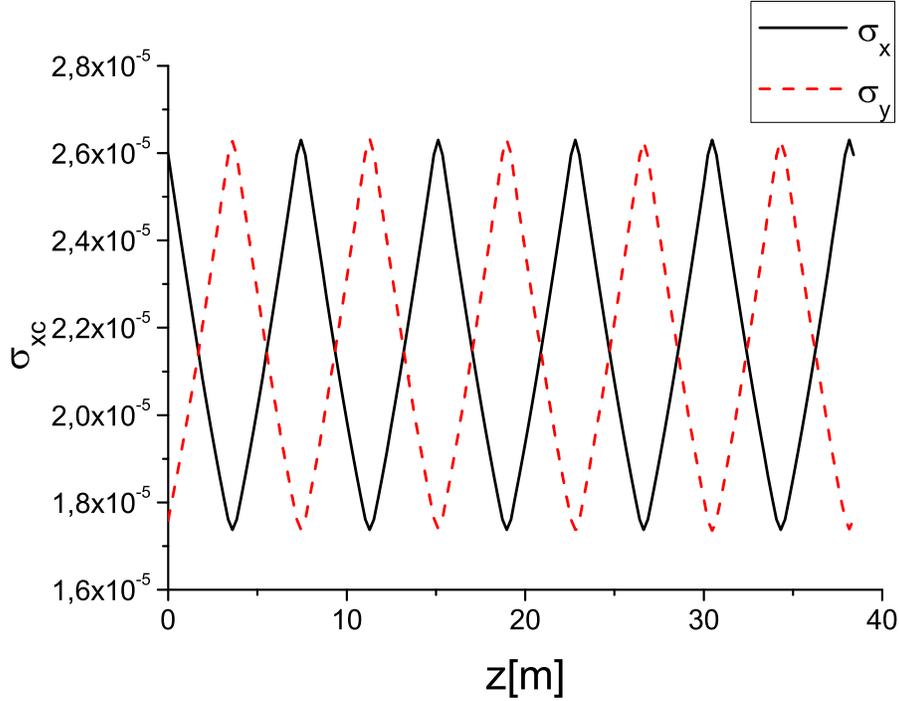}
\caption{Evolution of the rms horizontal and vertical beam size as a
function of the distance along the setup calculated through Genesis
at a normalized emittance $\varepsilon_n = 4 \cdot 10^7$ m and at an
electron beam energy of $4.3$ GeV ($\gamma = 8416$). }
\label{sigxysig}
\end{figure}

\begin{figure}[tb]
\includegraphics[width=1.0\textwidth]{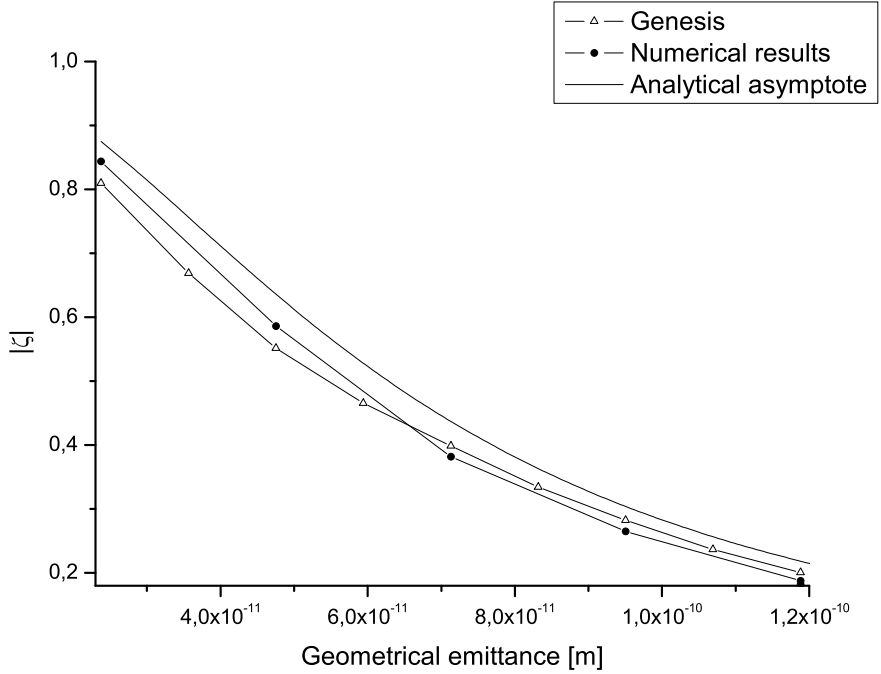}
\caption{Comparison between $|\zeta|$ calculated analytically as a
function of the geometrical emittance from Eq. (\ref{main3}) (solid
line), $|\zeta|$ calculated numerically (black circles) as described
in this Section and $|\zeta|$ calculated through Genesis.  }
\label{compar}
\end{figure}

Finally, we present a comparison between $|\zeta|$ obtained from Eq.
(\ref{main3}), calculated numerically as above, and derived using
the FEL code Genesis \cite{GENE}.  We assumed a cell length of
$3.84$ m, so that we set $L_\mathrm{FODO} = 3.84$m, and a drift
distance equivalent to $10$ cells, i.e. $L=38.4$ m. We set both
horizontal and vertical average betatron function to $10.1$ m. In
order to simulate the focusing system in Genesis, without the
influence of the undulators we set the electron beam current to
zero, we switched off the undulator focusing, and we prepared a
Genesis particle file with a given density bunching at the
modulation wavelength $\lambda = 1.5$ nm, so that it was matched
with the FODO beam transport line. The electron beam energy was set
to $4.3$ GeV. All particles in the particle file were set with the
same energy:  as a result effects of the momentum compaction factor
were excluded. The beam was propagated through the setup. The
evolution of the rms horizontal and vertical size as a function of
the distance along the setup is shown in Fig. \ref{sigxysig}. At the
end of the setup, the final particle beam was extracted, allowing
for a comparison of the final bunching with respect to the initial
bunching. The debunching as a function of the geometrical emittance
is presented in Fig. \ref{compar}.

\section{\label{sec:conc} Conclusions}

In this paper we derived an analytical expression for the debunching
of an modulated electron beam through a FODO focusing structure. The
expression is very simple, and can be practically used for any value
in the parameter space.

\section{Acknowledgements}

We are grateful to Massimo Altarelli, Reinhard Brinkmann, Serguei
Molodtsov and Edgar Weckert for their support and their interest
during the compilation of this work.


\begin{thebibliography}{99}


\bibitem{LCLS2} P. Emma et al., Nature photonics doi:10.1038/nphoton.2010.176 (2010)

\bibitem{GENG} H.Geng, Y. Ding and Z. Huang, Nucl.Instr. and Meth. A 622 (2010)   276

\bibitem{KUSK} B. Kuske and J. Bahrdt, FEL Conference Proceedings 2008 p.348, "Tolerance studies for APPLE undulators in FEL facilities"

\bibitem{OURC} G. Geloni, V. Kocharyan and E. Saldin, "Circular polarization control for the LCLS
baseline in the soft X-ray regime", DESY 10-252 (2010).


\bibitem{SVEN} S. Reiche, Nucl. Instrum. Methods A, 445, 90 (2000)

\bibitem{GENE} S Reiche et al., Nucl. Instr. and Meth. A 429, 243 (1999).




\end{thebibliography}
\end{document}